\documentclass{emulateapj}
\submitted{}
\journalinfo{To appear in ApJL, V.~629}

\def \lowz{\hbox{$z$$\sim$0.3}}
\def \hiz{\hbox{$z$$\sim$0.5}}
\def \zeroz{\hbox{$z$$\sim$0.05}}
\def \BR0{\hbox{$(B-R_c)_0$}}

\def\gtapr {\lower .1ex\hbox{\rlap{\raise .6ex\hbox{\hskip .3ex
        {\ifmmode{\scriptscriptstyle >}\else
                {$\scriptscriptstyle >$}\fi}}}
        \kern -.4ex{\ifmmode{\scriptscriptstyle \sim}\else
                {$\scriptscriptstyle\sim$}\fi}}}

\lefthead{Yee et al.}
\righthead{The Dependence of Galaxy Colors}
\begin{document}

\title{The Dependence of Galaxy Colors on Luminosity and Environment
at $z\sim0.4$\footnotemark[1]}

\author{H.K.C. Yee\altaffilmark{2}, B.C. Hsieh\altaffilmark{3,4},
H. Lin\altaffilmark{5}, 
M.D. Gladders\altaffilmark{6}}

\altaffiltext{1}{Based on observations from the Canada-France-Hawaii
Telescope (CFHT), which is operated by the National Research Council of
Canada, le Centre Nationale de la Recherche Scientifique and the
University of Hawaii.}
\altaffiltext{2}{Department of Astronomy \& Astrophysics,
        University of Toronto, 60 St. George Street, Toronto, Ontario
        M5S 3H8, Canada; email: hyee@astro.utoronto.ca}
\altaffiltext{3}{Institute of Astronomy, National Central University,
No. 300, Jhongda Rd. Jhongli City, Taoyuan County 320, Taiwan, R.O.C.}
\altaffiltext{4}{Institute of Astrophysics \& Astronomy, Academia Sinica,
P.O. Box 23-141, Taipei 106, Taiwan, R.O.C; email: 
paul@cluster.asiaa.sinica.edu.tw}
\altaffiltext{5}{Fermi National Accelerator Laboratory, P.O. Box 500, Batavia,
IL 60510; email: hlin@fnal.gov}
\altaffiltext{6}{Carnegie Observatories, Pasadena, CA 91101, USA;
email: gladders@ociw.edu}

\begin{abstract}
We analyse the $B-R_c$ colors of galaxies as functions of luminosity
and local galaxy density using a large photometric 
redshift catalog based on the Red-Sequence Cluster Survey.
We select two samples of galaxies with a magnitude limit of
$M_{R_c}<-18.5$ and redshift ranges of $0.2\le z < 0.4$ 
and $0.4\le z <0.6$
containing $\sim10^5$ galaxies each.
We model the color distributions of subsamples of galaxies
and derive the red galaxy
fraction and peak colors of red and blue galaxies as functions of galaxy 
luminosity and environment.
The evolution of these relationships over the redshift range
of \hiz~to \zeroz~is analysed in combination with published
results from the Sloan Digital Sky Survey.
We find that there is a strong evolution in the restframe peak color of
bright blue galaxies in that they become redder with decreasing redshift,
while the colors of faint blue galaxies remain approximately constant.
This effect supports the ``downsizing'' scenario
of star formation in galaxies.
While the general dependence of the galaxy color distributions 
on the environment is small, we find that
the change of red galaxy fraction with epoch is a function of
the local galaxy density, suggesting that the downsizing effect
may operate with different timescales in regions of different galaxy
densities.

\end{abstract}


\section{Introduction}

Many investigations in the past 30 years have shown 
that galaxy populations, manifested by galaxy colors, spectra, 
 and morphological distributions, have a strong
dependence on the environment and luminosity of the galaxies
(e.g., Melnick \& Sargent 1970; Dressler 1980; de Vaucouleurs 1961 and many
other subsequent studies).
More recently,  much more detailed analyzes of galaxy color
distributions in the local universe have
become available  using large galaxy samples of tens 
of thousands from the Sloan Digital Sky Survey (SDSS).
Baldry et al. (2004) and others show that the colors
of galaxies can be neatly separated into components of red and
blue galaxies.
Balogh et al. (2004, hereafter B04) 
were able to delineate the dependence of galaxy colors on galaxy
luminosity and environment, concluding that there is only a weak
 dependence on the latter.

In this Letter we present a similar study of galaxy samples at redshift
between 0.2 and 0.6 using the photometric redshift galaxy catalogs from
the Red-Sequence Cluster Survey (RCS) from Hsieh et al.~(2005, hereafter
H05). 
Combined with results from a \zeroz~sample from
the SDSS (B04), we  examine the trend of the dependence of 
galaxy colors on luminosity and environment at different epochs.
In \S2 we briefly describe the photometric redshift galaxy sample and the
measurement techniques.
We present the results in \S3;
their implications are discussed and summarized in \S4.
We adopt a flat cosmology with 
$\Omega_m=0.3$,
$\Omega_\Lambda=0.7$ and $H_0=70 $ km s$^{-1}$Mpc$^{-1}$.

\section{The Photometric Redshift Galaxy Data}

The RCS is a 92 square degree imaging survey in the $z'$ and  $R_c$ bands
conducted with the CFHT 3.6m and the CTIO 4m 
to search for galaxy clusters in the redshift range of
$z<1.4$ (see Gladders \& Yee 2005). 
Additional imaging in the $V$ and $B$ bands was also obtained 
for 33.6 square degrees in the northern patches using the CFH12K camera.
The observation and data reduction techniques are discussed in detail
in Gladders \& Yee (2005) for the $z'$ and $R_c$ data, and in H05  
for the $V$ and $B$ data.
Photometric redshifts for 1.2 million galaxies using the four-color
photometry are derived using an empirical training set method
with 4,924 spectroscopic redshifts.
Detailed descriptions of the photometric redshift 
method and the precision and completeness of the sample are presented in H05.

\begin{figure*}[!t]
\begin{center}
\leavevmode
\epsfxsize=6.0cm
\includegraphics[angle=0,width=0.83\textwidth]{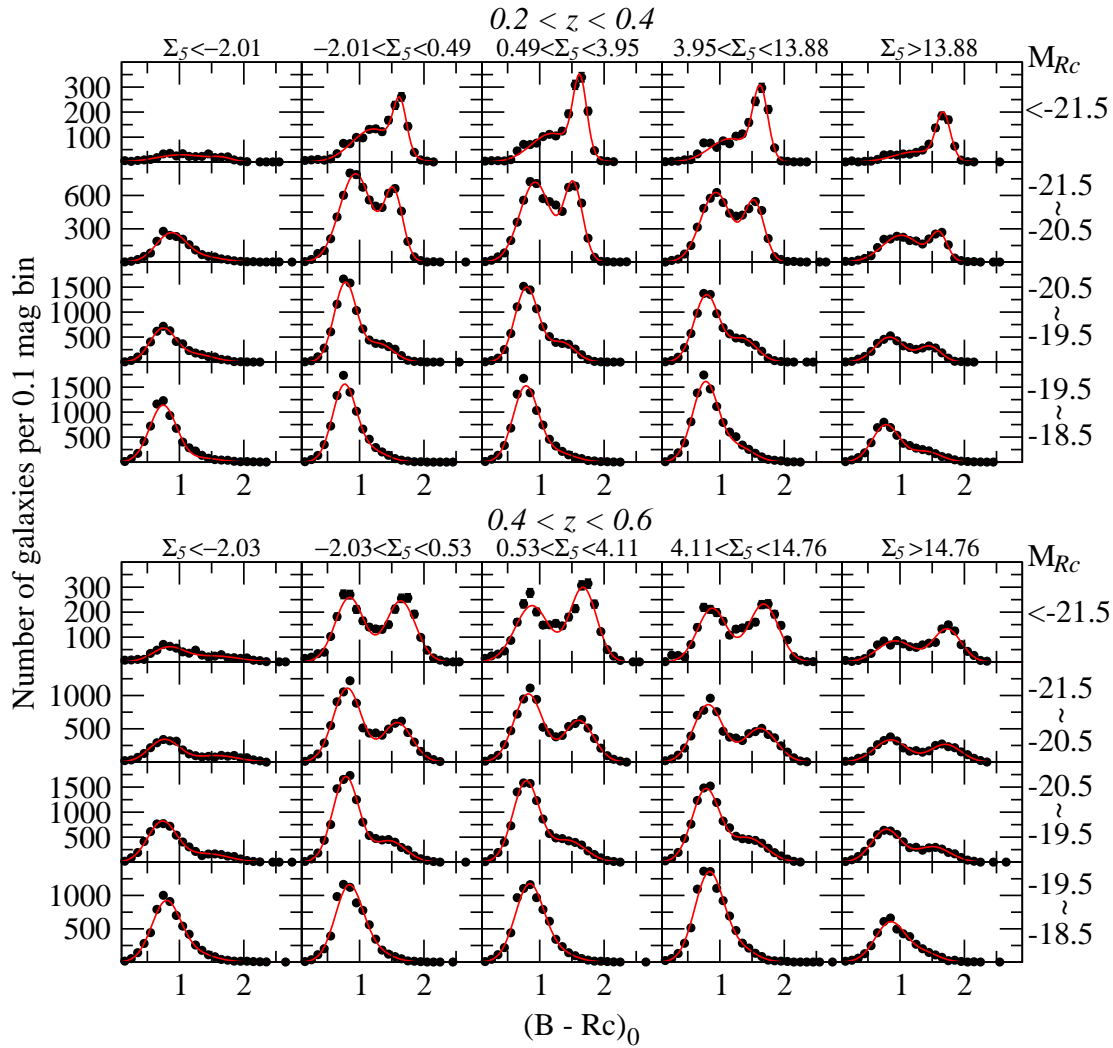}
\figcaption{\footnotesize {\it Upper large panel}: Distributions
of rest $(B-R)$ colors for galaxies in subsamples of different 
surface density ($\Sigma_5$) and luminosity ($M_{R_c}$) for
$0.2\le z<0.4$.
Subpanels for different densities are plotted horizontally,
while subpanels for different absolute magnitudes are plotted
vertically.
{\it Lower large panel}: same for $0.4\le z <0.6$. Solid lines
are two-Gaussian fits to the distributions of blue and red galaxies.
}
\label{colorhisto} 
\end{center}
\end{figure*}

We select samples of galaxies covering two moderate redshift intervals
using conservative criteria to minimize redshift and color errors.
We choose  galaxies with absolute magnitudes $M_{R_c}<-18.5$
in the photometric redshift bins: $0.2\le z<0.4$ (hereafter, the \lowz~sample)
and $0.4\le z<0.6$ (hereafter, the \hiz~sample) with
redshift uncertainty $\sigma_z/(1+z)<0.1$, where $\sigma_z$ is the
computed photometric redshift uncertainty (see H05).
The absolute magnitudes are corrected for K-correction and estimated
evolution.
For each galaxy, we use model colors computed from the
spectral energy distributions of
galaxies from Coleman et al.~(1980) to estimate a galaxy spectral type
based on the observed $R_c-z'$ color, and derive K-corrections for
the $B$ and $R_c$ magnitudes.
We approximate the $R_c$ band luminosity evolution by 
$M(z)=M(0)-Qz$ (see Lin et al. 1999),
where we adopted $Q=1.5$ for early-type galaxies ($B-R_c>1.8$) and
$Q=1.0$ for late-type galaxies ($B-R_c<1.8$).
We further limit our galaxy sample to be at least 100$''$ from the
edges of the fields.
These criteria produce a sample of 106,095 and 124,004
galaxies for the \lowz~and \hiz~sample, respectively.

We estimate the local galaxy density using a projected 
surface density as a proxy (see, e.g., Dressler 1980).
For each galaxy in the primary sample, we measure $R_5$, the distance to the
fifth nearest neighbor with $M_{R_c}\le-19.5$, from which a local 
surface density parameter $\Sigma_5$ in units of Mpc$^{-2}$ 
(proportional to $R_5^{-2}$) is computed.
For counting the five objects we select only those with photometric redshifts
within 2$\sigma_z$ of the primary galaxy.
A completeness correction for each neighbor is applied, as
not all galaxies within the sampling magnitude range 
satisfy the redshift uncertainty criterion. 
The completeness factor is estimated using the ratio of the total 
number of galaxies within 
a 0.1 mag bin at the magnitude of the neighbor galaxy to the number
of galaxies in that bin with photometric redshifts satisfying the
redshift uncertainty criterion.
Since red galaxies on average have lower photometric redshift
uncertainties, we refine our completeness correction by computing
the factor separately for red and blue galaxies,
separated at $B-R_c=1.8$.
The average completeness correction factor is $\sim$1.10.
We count the nearest neighbors to the primary galaxy by summing
the corrected weight for each galaxy until the total is $\ge5.0$.

To be able to put the $\Sigma_5$ parameter for galaxies at
different redshift on
a comparative scale, a background correction must be applied to $\Sigma_5$.
This is estimated for each primary galaxy
by computing the average number
of galaxies in a circle of radius $R_5$ using galaxies in the
same RCS patch (of size $\sim2.5$ deg $^2$, see Gladders \& Yee 2005)
that satisfy the redshift
and magnitude criteria used for the counting of the neighbors.
Completeness corrections are also applied in the counting of the
background galaxies.
The average background correction applied is 5.59 galaxies, with a scatter
of 4.43.
The stochastic nature of the background produces negative local
densities; for convenience, we add 5 to $\Sigma_5$ for the purpose
of displaying some of the results.

\begin{figure}
\leavevmode
\includegraphics[angle=-90,width=0.50\textwidth]{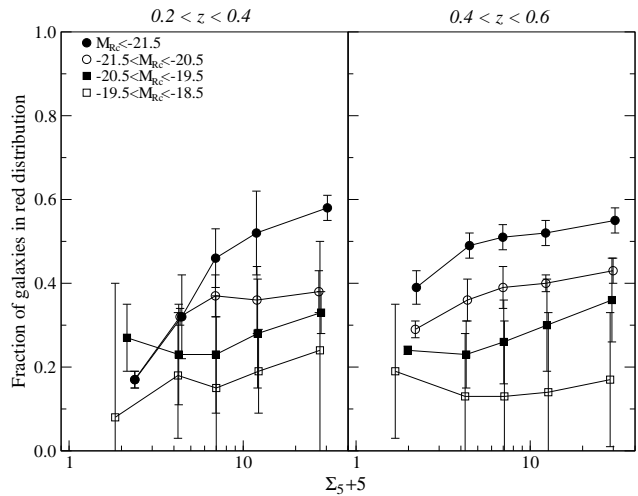}
\figcaption{\footnotesize Red galaxy fraction, $f_r$, as a function
of local galaxy density for samples of different luminosities
for the two redshift bins.
}
\label{redgalfrac} 
\end{figure}

\begin{figure}
\leavevmode
\epsfxsize=7.5cm
\includegraphics[angle=-90,width=0.50\textwidth]{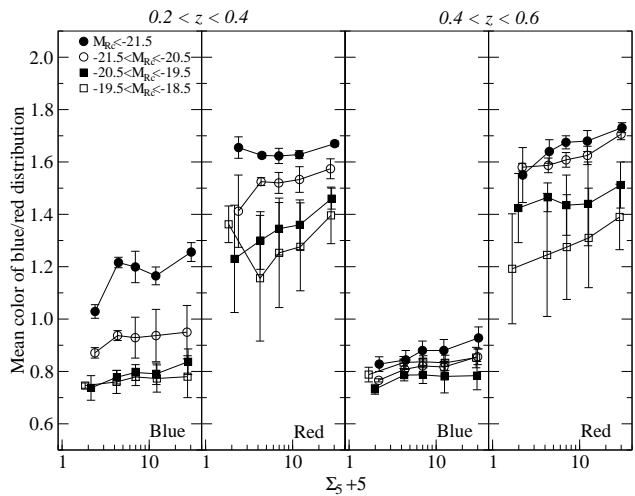}
\figcaption{\footnotesize Galaxy color (\BR0) peak of the blue and red galaxy
distributions as a function
of local galaxy density for samples of different luminosities
for the two redshift bins.
}
\label{colorpeak} 
\end{figure}
\section{Results}
The two panels of Figure~1
show the rest \BR0~distributions in bins of absolute 
magnitude and $\Sigma_5$ for the \lowz~and \hiz~bins.
We divide the galaxies into subsamples of local density such
that the first bin contains the least dense 12.5\% of the galaxies;
the next three bins, each of the successive 25\%; and the last bin,
the most dense 12.5\%.
The solid lines are the two-Gaussian fits for the red and blue populations.
Qualitatively, the two-Gaussian model that has been used for the SDSS
$z\sim0.05$ galaxy samples by Baldry et al. (2004) and B04 is also
a reasonable descriptor of the data at these higher redshifts.
For comparison purposes, we note that the approximate transformations
between the SDSS system and our data are:
$r\sim R_c+0.5$ and $u-r\sim 2(B-R_c)-0.4$
(based on galaxy color data from Fukugita et al.~1995).
We also repeat the calculation using a sample with a more
liberal error criterion, $\sigma_z/(1+z)<0.4$; this produces
qualitatively identical results, showing that the incompleteness
correction works well.

Figure 2 shows the fraction of red galaxies, $f_r$,  derived from 
integrating the 
Gaussian fits, as a function of local density for each magnitude
bin for the two redshift samples.
The uncertainty for each  $f_r$ value is estimated by
generating 100 Monte-Carlo
realizations based on the errors on the parameters of
the corresponding two-Gaussian fit. We take
the 68\% confidence limits on $f_r$ for the 100 results to be
the $\pm1\sigma$ uncertainty.
For both redshift bins, there is a strong dependence of $f_r$
on luminosity, covering a factor of 5--6,
with the luminous galaxy samples having the higher values.
Within each magnitude bin, the dependence of $f_r$ on local density,
however, is relatively moderate, especially for
the fainter galaxies.

Figure 3 illustrates the dependence of the
peak \BR0~of the Gaussian fits  on local density for galaxies
of different luminosities.
Similar to the SDSS result at \zeroz~(B04; Hogg et al.~2003), 
the peak colors of the color
distributions of both the red and the blue galaxies have only a
weak dependence on the local galaxy density.
We note that because of the scatter of the background corrections,
there will be some blurring of the local density parameter;
however, this smoothing should have minimal effect at the high density
bins, where the density is much higher than the scatter in the background
correction.

\section{Discussion and Conclusions }

Using the three redshift epochs, we find a number of interesting
evolutionary effects in the color distribution of galaxies.
We find the peak colors of the red galaxy distributions to be remarkably
similar over the three redshift bins. 
The color peaks for both the \lowz~and \hiz~samples for the different
magnitude subsamples
range from \BR0~of $\sim$ 1.3 
to $\sim1.6$, with the brighter red galaxies being redder.
This is essentially identical to the \zeroz~SDSS samples of B04 over 
a similar luminosity range, which have $u-r\sim2.2$ to 
2.5 (or $B-R_c\sim 1.3$ to 1.5).
Hence, there has been only minimal changes in the colors of 
red galaxies in all environments from \hiz~to \zeroz.
The small color changes suggest that red galaxies of all luminosities and in 
different environments are already well-evolved by redshift $\sim0.6$
(e.g., see models in Bruzual \& Charlot 1993).

The difference in \BR0~for red galaxies of different absolute magnitudes
is likely due to the slope of the color-magnitude relation (CMR) 
for early-type galaxies.
The range of \BR0~of $\sim$ 0.3 mag over $\sim4.5$ mag in $M_{R_c}$ is
equivalent to a CMR slope of $\sim$0.06, very similar to the slope
of 0.03 to 0.08 found
in the CMR of galaxy clusters (e.g., L\'opez-Cruz et al.~2004).

The blue galaxy color distributions, however, show a strong evolutionary
effect from \hiz~to $z$$\sim$0.
For the \hiz~sample, the peak colors of the blue galaxies are
essentially independent of the luminosity of the galaxies,
with values bunching up around \BR0~$\sim0.85$, with a dispersion
of less than 0.05 mag.
Thus, on average, for all magnitudes, blue galaxies are at least
as blue as the faintest blue galaxies at lower redshift.
For the \lowz~sample, the blue galaxies in the two brighter bins
($M_{R_c}<-20.5$) now have color peaks 0.2 to 0.4 mag redder than that
of the faintest blue galaxy bin.
Comparing to the \zeroz~sample of B04 (their Figure 3), the 
evolution is quite dramatic.
Here, the peak color of each successive brighter bin shifts significantly
to the red, to the point where the brightest blue
galaxy bin  has almost the same
color peak as the faintest red galaxy bin.
We also note that the bluest peaks in our data are about 0.1
mag bluer than the corresponding bin in B04.

This evolution of the blue galaxy peak colors can be interpreted
in the context of a `down-sizing' scenario, a term first suggested
by Cowie et al.~(1996).
To first order, we can assume that \BR0~provides a measure of the
specific star formation rate (SSFR,
defined as the star formation rate per unit mass), or the time
since the last major star formation episode 
(e.g., Bruzual \& Charlot 1993).
At \hiz, blue galaxies of all luminosities have very similar 
\BR0~color peaks, which is consistent with
galaxies of different masses having similar SSFRs.
The mean colors of the brighter blue galaxies
then  become progressively redder with time, while those of the faint galaxies
remain blue. 
This can be interpreted as the decrease in the SSFR being a
strong function of the luminosity of the galaxy: the more massive the galaxy,
the larger the decrease, or alternatively, the earlier in time
the decrease takes place.

This interpretation is in excellent agreement with the results of
Heavens et al.~(2004), who analysed the \zeroz~SDSS spectra
to recover the star formation history.
Their Figure 2 shows that low-mass galaxies have a more or less constant
star formation rate from $z\sim1$, peaking at $z\sim$ 0.4 to 0.2.
In comparison, the more massive galaxies have their star formation peak
much earlier, and show a decline as early as $z\sim1$,
and by \zeroz~the SSFRs for galaxies of different masses are different by a
factor of more than 10.
We note the complementary nature of the techniques used to arrive 
at a similar conclusion, in that
the Heavens et al.~work is based on the reconstruction of star formation 
histories from present-day stellar populations, whereas our results 
are derived from direct measurements of galaxy colors at
earlier epochs.

We find clear indications of evolution in $f_r$ from \hiz~to 
\zeroz, in that there is a general increase in $f_r$ at lower redshift. 
Comparing our high
redshift bins with the SDSS results (B04),
the evolution of $f_r$ also appears to be a function of the
local galaxy density: the \hiz~sample
has a much flatter dependence of $f_r$ on local density.
In B04's Figure 2 there is a strong local density dependence
of $f_r$ for galaxies of {\it all} luminosities; $f_r$ ranges
from 0.1 to 0.4 for galaxies of different luminosities in the lowest
density sample, and from 0.65 to 0.85 for the highest density samples.
By comparison, $f_r$ in our two higher redshift bins has a similar
range (0.1 to 0.4) for the low density environment, but has values
between 0.15 and 0.6 in the densest regions.
Thus, there is a lack
of faint red galaxies in the high density bins:
$f_r<0.4$ for all but the most luminous galaxies.
In other words, at $z\gtapr0.2$, there are many moderately luminous to
faint blue galaxies in high density regions.
This drop of $f_r$ at high density and high redshift
could be a reflection of the change in the morphology-density
relation at $z\sim0.5$ (e.g., Dressler et al.~1997; Smith et al.~2005).
The lack of faint red galaxies in high density regions
has also been noted by others at moderate to high redshifts
(e.g., Kodama et al.~2004)


The density-dependent evolution of $f_r$  suggests
the need for an additional element in the  downsizing scenario:
the downsizing timescale is a function of the environment.
That is, for galaxies of similar masses, the drop in star formation
rate occurs successively later for galaxies in less dense environments.
The increase of the faint blue galaxy fraction in the  high redshift,
high density samples 
can be interpreted as these galaxies having their
peak star formation era at a
redshift of $\sim0.3$ to 0.2 (Heavens et al.~2004), followed by a decline,
resulting in a high faint red galaxy fraction at \zeroz~(Figure 2, B04).
In the less dense regions, the star formation rate continues to be high
to lower redshifts (see, e.g., G\'omez et al.~2003), 
keeping the blue galaxy fraction high.
This is supported qualitatively by the positive, if small, gradients in the
\BR0~peak going from the least to the most dense samples,
for all redshift bins.

In summary we have analysed the \BR0~color distribution of 
a sample of $\sim230,000$ galaxies with photometric redshifts
between 0.2 and 0.6 and $M_{R_c}<-18.5$, in combination with
the results derived at \zeroz~by B04 using SDSS data.
We find there is a strong evolution in the peak \BR0~colors
of the blue galaxies, consistent with a downsizing
star formation scenario where more massive galaxies have their
peak star formation time occurring at earlier epochs.
This is in excellent qualitative agreement with
star formation histories of galaxies of different masses
derived by Heavens et al. (2004) using the technique of
``fossil stellar populations.''
We also find that the evolution of the red galaxy fraction
is dependent on local galaxy density, in that there is a 
significant deficit of faint red galaxies in rich environments at
high redshift.
In a future paper we will examine the evolution of the color
distributions for galaxies in different local densities in
various global environments (e.g., cluster core, cluster outskirt, 
and field), and
interpret the results more quantitatively in terms of stellar
mass and star formation rate via stellar population models.

\end{document}